\documentclass[twocolumn,showpacs,showkeys,preprintnumbers,amsmath,amssymb]{revtex4}

\usepackage{graphicx}
\usepackage{dcolumn}
\usepackage{bm}
\usepackage{enumerate}

\begin{document}

\preprint{To be published in Phys. Rev. B}

\title{Defect healing at room temperature in pentacene thin films \\ and improved transistor performance}

\author{Wolfgang L. Kalb}%
 \email{kalb@phys.ethz.ch}
\author{Fabian Meier}
\author{Kurt Mattenberger}
\author{Bertram Batlogg}
\affiliation{%
Laboratory for Solid State Physics, ETH Zurich, 8093 Zurich,
Switzerland
}%

\date{\today}

\begin{abstract}
We report on a healing of defects at room temperature in the organic
semiconductor pentacene. This peculiar effect is a direct
consequence of the weak intermolecular interaction which is
characteristic of organic semiconductors. Pentacene thin-film
transistors were fabricated and characterized by in situ gated
four-terminal measurements. Under high vacuum conditions (base
pressure of order 10$^{-8}$\,mbar), the device performance is found
to improve with time. The effective field-effect mobility increases
by as much as a factor of two and mobilities up to 0.45\,cm$^{2}$/Vs
were achieved. In addition, the contact resistance decreases by more
than an order of magnitude and there is a significant reduction in
current hysteresis. Oxygen/nitrogen exposure and annealing
experiments show the improvement of the electronic parameters to be
driven by a thermally promoted process and not by chemical doping.
In order to extract the spectral density of trap states from the
transistor characteristics, we have implemented a powerful scheme
which allows for a calculation of the trap densities with high
accuracy in a straightforward fashion. We show the performance
improvement to be due to a reduction in the density of shallow traps
$\leq0.15$\,eV from the valence band edge, while the energetically
deeper traps are essentially unaffected. This work contributes to an
understanding of the shallow traps in organic semiconductors and
identifies structural point defects within the grains of the
polycrystalline thin films as a major cause.
\end{abstract}

\maketitle

\section{\label{sec:level1} Introduction}

Organic electronics now rapidly enters the consumer markets.
Field-effect mobilities in polycrystalline organic semiconductors
are comparable to the mobilities in hydrogenated amorphous silicon
and the organic semiconductors can be deposited by thermal
evaporation or from solution at low costs on large areas. Pentacene
and its soluble derivatives are among the most promising organic
semiconductors for organic thin-film transistor
applications.\cite{LinYY1997, SherawCD2003}

Organic semiconductors are distinct in that they generally consist
of neutral molecules which interact by rather weak forces
(predominantly Van der Waals forces). The energy associated with a
Van der Waals bond is 10$^{-3}$-10$^{-2}$\,eV, i.e. orders of
magnitude smaller than the energy of a covalent bond (several
eV).\cite{SilinshEA1994} Therefore, organic semiconductors are soft,
have a low mechanic strength and a high compressibility,  low
melting and sublimation temperatures and a large thermal expansion
coefficient. The weak interaction forces between the molecules not
only enable low-cost processing but also lead to peculiarities such
as the existence of at least four different pentacene polymorphs or
an even negative thermal expansion coefficient.\cite{MatheusCC2001,
HaasS2007}

As in the case of amorphous inorganic materials, the charge
transport in organic semiconductors can be described with a mobility
edge or transport level concept. In the case of crystalline small
molecule organic semiconductors such as pentacene, the mobility edge
may be identified with the valence or conduction band edge and the
charge carriers are transported in extended states. However, the
charge carriers are transported in states above a mobility edge and
are multiply trapped by and thermally released from trap states
below the mobility edge. Consequently, the conduction of charge in
organic semiconductors and thus the transistor characteristics
critically depend on trap states. Energetically deep traps influence
the subthreshold swing of a device which needs to be sufficiently
steep for a low power operation. The density of shallow traps, on
the contrary, can dominate the effective field-effect mobility. Trap
states in organic semiconductors have long been a subject of
investigation.\cite{ProbstKH1975} This topic is currently attracting
much attention due to the crucial importance of trap states for the
emerging applications of organic field-effect
transistors.\cite{NorthrupJE2003, JurchescuOD2004, KangJH2005,
GoldmannC2006, KalbWL2007, KrellnerC2007} The shallow traps in
particular are poorly understood.

The density of trap states as a function of energy can be derived
from the transistor characteristics. On the one hand, a density of
states function can be postulated a priori and the corresponding
transistor characteristic can be simulated by means of a suitable
computer program. The density of states function is then iteratively
refined until, after a number of predictor-corrector loops, good
agreement between the measured characteristic and the simulated
curve is achieved.\cite{VoelkelAR2002, ScheinertS2002,
OberhoffD2007} On the other hand, the density of states function can
be calculated from the linear regime transfer characteristics in a
straightforward fashion. A number of such extraction schemes has
been suggested.\cite{HorowitzG1995, SchauerF1999, LangDV2004,
DeAngelisF2006} This approach has the advantage of giving an
unambiguous result but, depending on the complexity of the
extraction scheme, spurious errors may result from simplifying
assumptions.

Organic thin-film transistors (TFTs) are most often characterized
after the samples have been exposed to ambient air and in situ
electrical characterization is very rare.\cite{KiguchiM2005,
KnippD2007} The organic semiconductors, however, are generally
presumed to be sensitive to water vapor and oxygen. Moreover, the
transistors are often characterized by two-terminal measurements
which do not allow to distinguish between contact effects and
effects of the semiconducting layer. The gated four-terminal method
yields the field-effect conductivity and the effective field-effect
mobility free from contact effects. It also allows for an extraction
of the device contact resistance.\cite{TakeyaJ2003, GoldmannC2004,
PesaventoPV2004, PesaventoPV2006}

For this work pentacene served as a prototypical small molecule
semiconductor and we have investigated the performance of pentacene
thin-film transistors by in situ gated four-terminal measurements.
The corresponding trap states functions were derived with a
straightforward extraction scheme which had been successfully used
to understand trap states in hydrogenated amorphous silicon.

\section{Experimental}

\subsection{Device fabrication}

\begin{figure}
\includegraphics[width=0.95\linewidth]{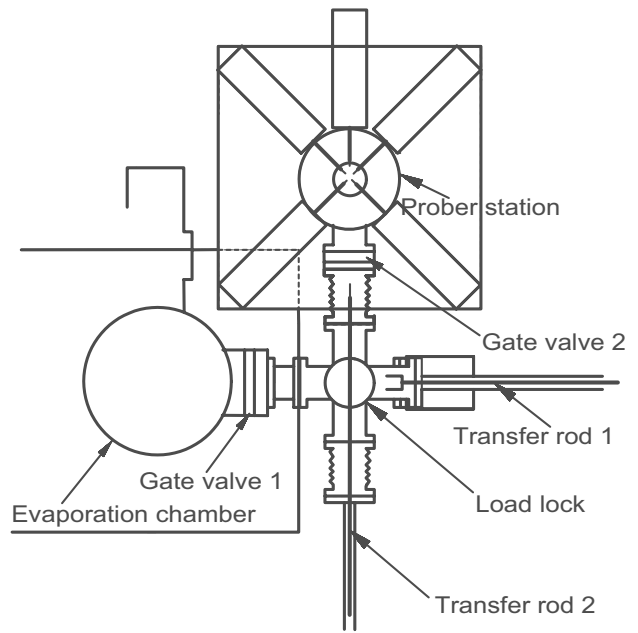}
\caption{\label{insitu} In situ device fabrication and
characterization system: pentacene thin-film transistors were
fabricated and characterized electrically by gated four-terminal
measurements at a base pressure of order 10$^{-8}$\,mbar without
breaking the high vacuum.}
\end{figure}

Pentacene from Aldrich (purum) was sublimation purified twice and
was introduced into the evaporation chamber soon after the
purification. As substrates we used heavily doped Si wafers with a
260\,nm thick SiO$_{2}$ layer. The substrates were cleaned with hot
acetone and hot isopropanol (MOS grade) in an ultrasonic bath.
Immediately after the cleaning, the substrates were mounted on a
sample holder and were introduced into the in situ device
fabrication and characterization system via the load lock
(Fig.~\ref{insitu}). The evaporation chamber and the prober station
were both separated from the load lock with a gate valve and were
constantly kept under vacuum (base pressure
$\approx3\times10^{-8}$\,mbar). The vacuum in the evaporation
chamber and in the prober station was maintained with a cryopump and
a turbopump, respectively. The substrates were introduced into the
evaporation chamber with transfer rod 1 and were placed on a shadow
mask for the pentacene evaporation. A high precision mask
positioning mechanism allowed for a proper adjustment of the mask
with respect to the substrates.

The substrates were kept in high vacuum for approximately 24\,h
prior to the device fabrication. After that time, also the pressure
in the turbo-pumped load lock was $\approx3\times10^{-8}$\,mbar. A
50\,nm thick film of pentacene was evaporated through the shadow
mask onto the Si/SiO$_{2}$ substrates while the substrates were kept
at room temperature. After the pentacene evaporation, the sample
holder was positioned on a shadow mask for the gold evaporation with
transfer rod 1 and the pentacene TFTs were completed by evaporating
40\,nm thick gold electrodes.

\begin{figure}
\includegraphics[width=0.5\linewidth]{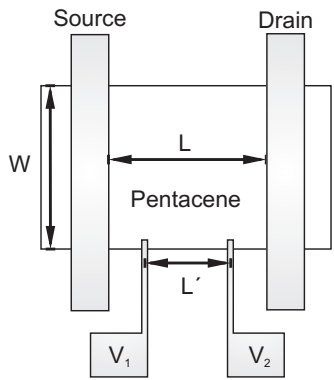}
\caption{\label{device} Transistor test structure for the gated
four-terminal measurements. The alignment was achieved by means of a
high precision in situ mask positioning mechanism. The channel
length and width were $L=450$\,$\mu$m and $W=1000$\,$\mu$m and the
distance between the voltage sensing electrodes was
$L'=300$\,$\mu$m. With the voltage sensing electrodes, the
potentials $V_{1}$ and $V_{2}$ were measured.}
\end{figure}

The resulting transistor test structures are shown schematically in
Fig.~\ref{device}. The transistors consisted of a well-defined
stripe of pentacene and had voltage sensing electrodes with little
overlap to the pentacene film. It has been shown that the use of a
``masked'' pentacene film and a proper alignment of the electrodes
is important for the four-terminal
measurement.\cite{PesaventoPV2006} The channel length and width of
the devices was $L=450$\,$\mu$m and $W=1000$\,$\mu$m, respectively.
The voltage sensing electrodes were situated at $(1/6)L$ and
$(5/6)L$, such that the distance between these electrodes was
$L'=300$\,$\mu$m.

After the completion of the device fabrication, the samples were
transported to the prober station by means of transfer rod 1 and
transfer rod 2 (Fig.~\ref{insitu}).

\subsection{Electrical characterization}

For the electrical measurements we used a HP 4155A semiconductor
parameter analyzer connected to five microprobers at the prober
station. Transfer characteristics in the linear regime were measured
in steps of 0.2\,V (drain voltage $V_{d}=-2$\,V). In all cases, the
forward and the reverse sweeps were measured. The integration time
was 20\,ms and the delay time was 0\,s. In addition to the drain
current $I_{d}$, the voltage drops $V_{1}$ and $V_{2}$ between the
grounded source and the two voltage sensing electrodes were measured
at each gate voltage $V_{g}$ (gated four-terminal measurement). All
electrical measurements were carried out in the dark.

A device was measured initially $\approx4$\,h after the completion
of the pentacene evaporation. Subsequently, the same device was
measured regularly (normally twice a day) for approximately one week
while being kept in the prober station at
$\approx3\times10^{-8}$\,mbar in the dark.

In some experiments, the effect of oxygen or nitrogen exposure on
the device characteristics was investigated. This was done by
introducing a continuous flow of gas (pureness
$\geq99.9999$\,Vol.-$\%$) into the prober station through a leak
valve, thus adjusting the partial pressure of the gas within the
prober station. A few days prior to the device fabrication for the
gas exposure experiments, the prober station was filled with the
respective gas through the leak valve in order to flush the gas
supply line and the valve. The prober station was then re-evacuated
and the supply lines were held at an overpressure of $0.5$\,atm
until the leak valve was opened in the experiment.

In other experiments, the influence of annealing was explored by
means of an electrical heating element on the sample holder tray in
the prober station.

\section{Parameter extraction}

In this section, we elucidate the extraction of key parameters from
the device characteristics. Linear regime transfer characteristics
are particularly interesting for the study of semiconductor physics,
because the gradual channel approximation is valid which leads to a
drastic simplification of the device physics. The two-dimensional
problem is approximated by two one-dimensional equations, i.e.
Possion's equation for the charge distribution perpendicular to the
insulator-semiconductor interface and a simple ohmic current-voltage
relationship.

\subsection{Basic parameter extraction}

In single crystalline inorganic MOSFETs above a threshold voltage,
essentially all the trap states are filled and the charge which is
induced by the gate appears in the valence/conduction
band.\cite{ShurM1984} Both the constant mobility in the above
threshold regime and the threshold voltage are important device
parameters. This approach may also be valid in the case of organic
single crystal transistors, i.e. organic field-effect transistors
with a very low trap density. However, contact effects often
severely affect the characteristics of these devices because of the
low channel resistance.\cite{PodzorovV2003}

As in the case of amorphous silicon field-effect transistors, the
equations developed for single crystalline MOSFETs are not suitable
to describe organic TFTs due to the increased trap
density.\cite{ShurM1984, HorowitzG1999, HorowitzG2004} Depending on
the density of trap states, the majority of charge carriers induced
by the gate may be trapped even at relatively high gate
voltages.\cite{OberhoffD2007} Provided that contact effects are
negligible, the drain current of an organic TFT in the linear regime
is given by
\begin{equation}
I_{d}=(W/L)\sigma V_{d}.
\end{equation}
$\sigma$ is the field-effect conductivity which is the effective
field-effect mobility $\mu_{eff}$ multiplied with the total gate
induced charge per unit area $C_{i}(V_{g}-V_{FB})$, i.e
\begin{equation}\label{sigma1}
\sigma=\mu_{eff}C_{i}(V_{g}-V_{FB}).
\end{equation}
$V_{FB}$ is the flatband voltage and $C_{i}$ the capacitance per
unit area. The effective field-effect mobility is one of the most
important device parameters and, for a p-type semiconductor such as
pentacene, $\mu_{eff}$ can be written as
\begin{equation}\label{mu0}
\mu_{eff}=\frac{P_{free}}{P_{free}+P_{trapped}}\mu_{0},
\end{equation}
where $P_{free}$ and $P_{trapped}$ are respectively the density of
free and trapped holes per unit area. $\mu_{0}$ is the extended
state mobility. The effective field-effect mobility $\mu_{eff}$ is
expected to increase with gate voltage even at relatively high gate
voltages, because the ratio $P_{free}/(P_{free}+P_{trapped})$
increases as the valence band is bent towards the Fermi
energy.\cite{HorowitzG1999}

The flatband voltage is the gate voltage which needs to be applied
in order to enforce flat bands at the insulator-semiconductor
interface. A non-zero flatband voltage can result from a difference
of the semiconductor and the gate Fermi levels. More importantly,
the flatband voltage is influenced by charge that is permanently
trapped at the interface or within the gate dielectric and only the
latter contribution can explain the large shifts of the transfer
characteristics that are sometimes observed in
experiments.\cite{WangA2006} The flatband voltage is approximately
equal to the onset voltage of the device. The onset voltage is
defined as the gate voltage, where the drain current, as a function
of gate voltage, rises sharply if plotted on a logarithmic scale. In
the present work we assume that the flatband voltage is equal to the
onset voltage.

From a technological point of view it is useful to define a
threshold voltage which marks the transition between the regime
below threshold to the regime above threshold. In the
above-threshold regime, the deep traps are filled and the
field-effect mobility is less strongly dependent on gate voltage.
The above-threshold regime in an organic TFT can be understood as
being in between the below-threshold regime and the above-threshold
regime of a single crystalline MOSFET.\cite{ShurM1984} It can be
shown that the field-effect mobility in the above-threshold regime
follows a power law $\mu_{eff}=\kappa(V_{g}-V_{t})^{\alpha}$ and
this allows for a refined extraction of the field-effect mobility
and the threshold voltage $V_{t}$.\cite{ShurM1984, HorowitzG2000,
HorowitzG2004}

In this study we do, however, follow the IEEE Standard for the
characterization of organic transistors and materials. The
field-effect conductivity can be calculated from
\begin{equation} \label{sigma0}
\sigma(V_{g})=\frac{L}{W}\frac{I_{d}}{V_{d}}.
\end{equation}
With Eq.~\ref{sigma1} and Eq.~\ref{sigma0} $\mu_{eff}$ can be
approximated as
\begin{equation} \label{mum1}
\mu_{eff}(V_{g})=\frac{1}{C_{i}}\bigg(\frac{\partial\sigma}{\partial
V_{g}}\bigg)_{V_{d}}=\frac{L}{WV_{d}C_{i}}\bigg(\frac{\partial
I_{d}}{\partial V_{g}}\bigg)_{V_{d}}.
\end{equation}
Since this approach is frequently used it has the advantage that the
values of $\mu_{eff}$ can easily be compared. Moreover, the
definition and extraction of a threshold voltage is not necessary.
However, small errors in the absolute value of the field-effect
mobility and its gate voltage dependence are to be expected since
the derivation rests on the assumption of a weak dependence of the
field-effect mobility on gate voltage.\cite{HorowitzG2004}

In an organic field-effect transistor, a significant fraction of the
drain voltage $V_{d}$ may drop at the contacts which can introduce
significant errors when extracting the field-effect conductivity and
the field-effect mobility. From gated-four terminal measurements,
the conductivity can be derived without error with
$I_{d}=(W/L')\sigma V_{d}'$. $L'$ is the distance between the
voltage sensing electrodes and $V_{d}'=V_{1}-V_{2}$ the voltage drop
between these electrodes (Fig.~\ref{device}). The contact-corrected
field-effect conductivity is then given by
\begin{equation}\label{sigma2}
\sigma(V_{g})=\frac{L'}{W}\frac{I_{d}}{(V_{1}-V_{2})}.
\end{equation}
The effective field-effect mobility $\mu_{eff}$ is not influenced by
contact effects when calculated from
\begin{equation}\label{mu}
\mu_{eff}(V_{g})=\frac{L'}{W(V_{1}-V_{2})C_{i}}\bigg(\frac{\partial
I_{d}}{\partial V_{g}}\bigg)_{V_{d}}.
\end{equation}

In the following we use the expressions ``two-terminal
conductivity'' and ``two-terminal mobility'' as short hand for
Eq.~\ref{sigma0} and Eq.~\ref{mum1}. We furthermore use the
abbreviations ``four-terminal conductivity'' and ``four-terminal
mobility'' for the quantities defined in Eq.~\ref{sigma2} and
Eq.~\ref{mu}.

The device contact resistance $R_{contact}$ was extracted from the
four-terminal measurement and was compared to the device channel
resistance $R_{channel}$. We now assume a linear voltage drop all
along the channel, i.e. from the source to the drain. With this
assumption, the voltage drop across the transistor channel is
$(V_{1}-V_{2})L/L'$ and the voltage drop at the contacts is
$V_{d}-(V_{1}-V_{2})L/L'$. The contact resistance is thus given by
\begin{equation} \label{rcontact}
R_{contact}(V_{g})=\frac{V_{d}-(V_{1}-V_{2})L/L'}{I_{d}}
\end{equation}
and the channel resistance by
\begin{equation} \label{rchannel}
R_{channel}(V_{g})=\frac{(V_{1}-V_{2})L/L'}{I_{d}}.
\end{equation}

\subsection{\label{advanced}Advanced parameter extraction}

The gate voltage dependence of the field-effect mobility reflects
the spectral density of trap states close to the valence band. A
number of direct extraction schemes has been suggested to obtain the
underlying density of states function.\cite{HorowitzG1995,
SchauerF1999, LangDV2004, DeAngelisF2006} In these approaches, the
relevant energy scale is derived from the activation energy
$E_{a}(V_{g})$ of the current (i.e. the field-effect conductivity)
which is obtained from temperature dependent measurements. If,
however, the electrical characteristics of a transistor change on a
timescale comparable to the time of a temperature dependent
measurement (hours), this approach is not suitable.

Gr\"unewald et al. have suggested an extraction scheme of high
accuracy for amorphous silicon thin-film transistors which allows to
convert a single linear regime transfer characteristic into the
underlying density of states function.\cite{GrunewaldM1980,
GrunewaldM1981, WeberK1982} It is based on surprisingly few
simplifying assumptions including:
\begin{enumerate}[a)]
\item the gradual channel approximation,
\item the semiconductor is homogenous normal to the
dielectric-semiconductor interface, and
\item insulator surface states only introduce an initial band
bending without applied field, i.e. contribute to a non-zero
flatband voltage $V_{FB}$.
\end{enumerate}
Extraction schemes often rest on the abrupt approximation: all the
charge is assumed to reside in a region close to the
dielectric-semiconductor interface of depth $\lambda(V_{g})$.
Gr\"unewald's method, however, is not based on this simplification
but takes proper account of the gate-induced band bending. In the
following, the key equations of the extraction scheme are given.

\begin{figure}
\includegraphics[width=0.45\linewidth]{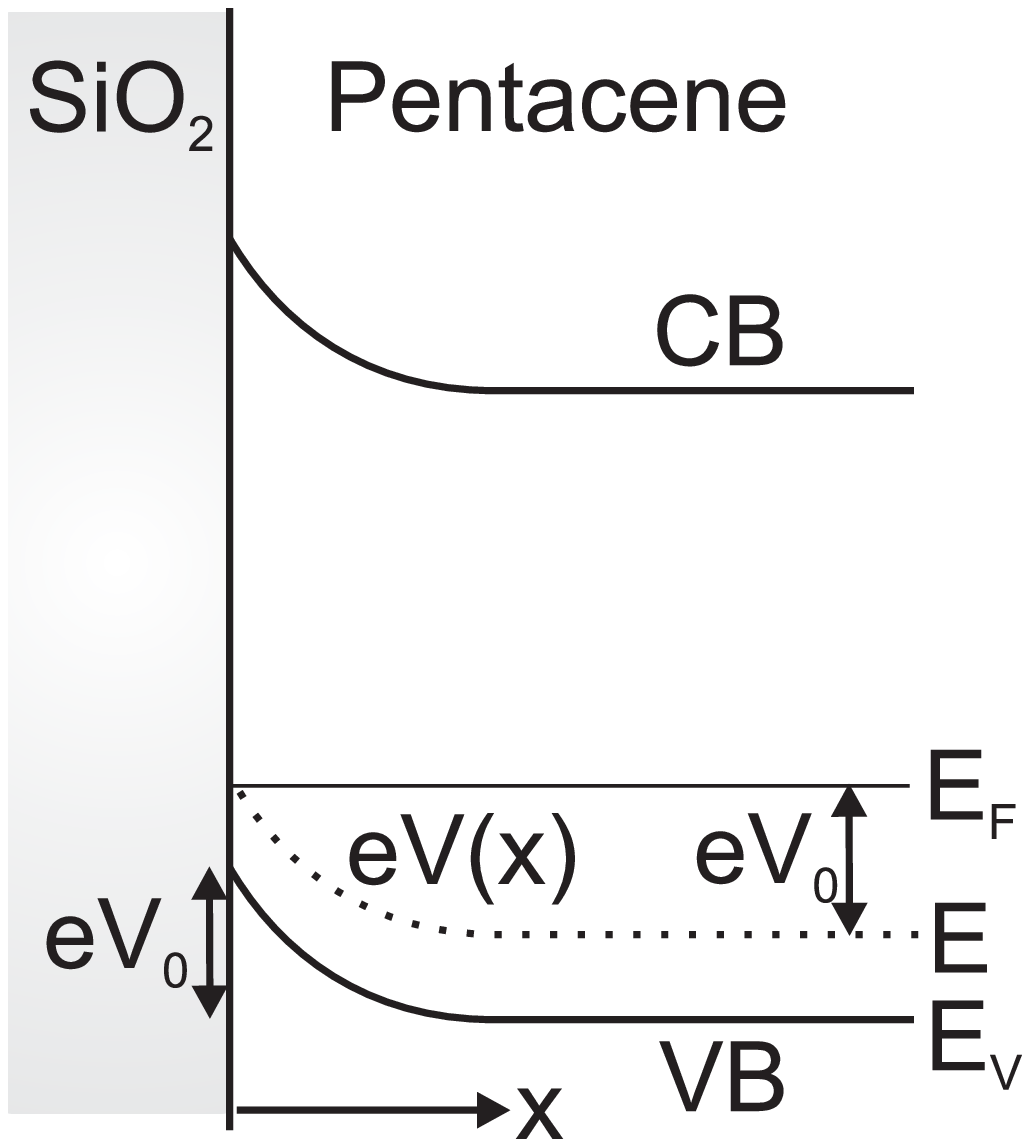}
\caption{\label{bandbending} Sketch of the energetics near the
SiO$_{2}$/pentacene interface: the application of a gate voltage
$V_{g}$ above the flatband voltage leads to a bending of the valence
band (VB) and of the conduction band (CB). At the interface ($x=0$),
the band shift is $eV(x=0)=eV_{0}$. Under these conditions, the
energy $E$ of specific trap states (dashed line) is raised at the
interface such that it coincides with the Fermi energy $E_{F}$ of
the sample.}
\end{figure}

Within the Boltzmann approximation, the field-effect conductivity is
given by
\begin{equation} \label{sigma3}
\sigma(V_{g})=\frac{\sigma_{0}}{d}\int_{0}^{d}\exp\bigg(\frac{eV(x)}{kT}\bigg)dx.
\end{equation}
Now $V_{g}$ is the gate voltage above the flatband voltage, $d$ is
the thickness of the pentacene film, $eV(x)$ is the band shift as a
function of the distance $x$ from the insulator-semiconductor
interface and $\sigma_{0}=\sigma(V_{g}=0)$ is the conductivity at
flatband. $\sigma_{0}$ can be approximated as
\begin{equation} \label{sigmanull}
\sigma_{0}=e\mu_{0}d\,N_{V}\exp\bigg(-\frac{E_{V}-E_{F}}{kT}\bigg)=e\mu_{0}\,P_{free}.
\end{equation}
$N_{V}$ is an effective density of extended states and $E_{V}$ the
energy of the valence band edge far from the insulator-semiconductor
interface. The situation is depicted in Fig.~\ref{bandbending}. The
complicated dependence of the band shift on the space coordinate $x$
in Eq.~\ref{sigma3} can be eliminated and an equation can be
derived, which implicitly contains the interface potential
$V(x=0)=V_{0}$ as a function of gate voltage:
\begin{eqnarray} \label{implicit}
\exp \bigg(\frac{eV_{0}}{kT}\bigg)-\frac{eV_{0}}{kT}-1 \nonumber
\\
=\frac{e}{kT}\frac{\epsilon_{i}d}{\epsilon_{s}l\sigma_{0}}\bigg[V_{g}\sigma(V_{g})-\int_{0}^{V_{g}}\sigma(V_{g}')dV_{g}'\bigg]
\end{eqnarray}
$l$ is the thickness of the gate insulator and $\epsilon_{i}$ and
$\epsilon_{s}$ are the dielectric constants of the insulator and the
semiconductor, respectively (see Ref. \onlinecite{GrunewaldM1980}
for a derivation of this equation). For each gate voltage,
Eq.~\ref{implicit} can be evaluated numerically and a value for the
interface potential $V_{0}$ is obtained. Eventually, we have the
complete function $V_{0}=V_{0}(V_{g})$.

In most general terms, the total hole density
$p=p_{free}+p_{trapped}$ (volume density) is
\begin{equation} \label{total1}
p(V)=\int_{-\infty}^{+\infty}N(E)[f(E-eV)-f(E)]dE,
\end{equation}
where N(E) is the density of states function. The total hole density
$p$ can be calculated as
\begin{equation} \label{total2}
p(V_{0})=\frac{\epsilon_{0}\epsilon_{i}^{2}}{\epsilon_{s}l^{2}e}V_{g}\bigg(\frac{\partial
V_{0}}{\partial V_{g}}\bigg)^{-1},
\end{equation}
i.e. by a numerical differentiation of the previously obtained set
of data $V_{0}=V_{0}(V_{g})$.\cite{GrunewaldM1981} Differentiating
Eq.~\ref{total1} with respect to the band shift yields
\begin{equation}
\frac{1}{e}\frac{dp(V)}{dV}=\int_{-\infty}^{+\infty}N(E)\bigg|\frac{df(E-eV)}{d(E-eV)}\bigg|dE.
\end{equation}
Several deconvolution methods exist to solve this type of equation
for $N(E)$, e.g. with cubic spline functions.\cite{SchauerF1986,
KrellnerC2007} However, for a slowly varying density of states
function (absence of monoenergetic states), the difference between
$N(E)$ and $dp/edV$ is expected to be relatively small on a
logarithmic scale.\cite{SchauerF1986} Consequently, we obtain the
final result by a numerical derivation of $p(V_{0})$ with respect to
the interface potential $V_{0}$ according to
\begin{equation} \label{total3}
\frac{1}{e}\frac{dp(V_{0})}{dV_{0}}\approx N(E).
\end{equation}
Within this zero temperature approximation, the band shift at the
interface $eV_{0}$ is equal to the energy of the respective traps
relative to the Fermi energy $E_{F}$ of the sample, i.e.
$eV_{0}=E-E_{F}$ (Fig.~\ref{bandbending}).

We have used Gr\"unewald's method to interpret the current voltage
characteristics from pentacene TFTs. We used the ``four-terminal
conductivity'' as a starting point which allows to extract a density
of states function free from contact artifacts. A simple
MATLAB$^{\circledR}$ coding allowed for the calculation of the
density of states function from Eq.~\ref{implicit}, \ref{total2} and
\ref{total3}.

\section{Results}

\subsection{Improvement of the device performance with time}

\begin{figure}
\includegraphics[width=0.9\linewidth]{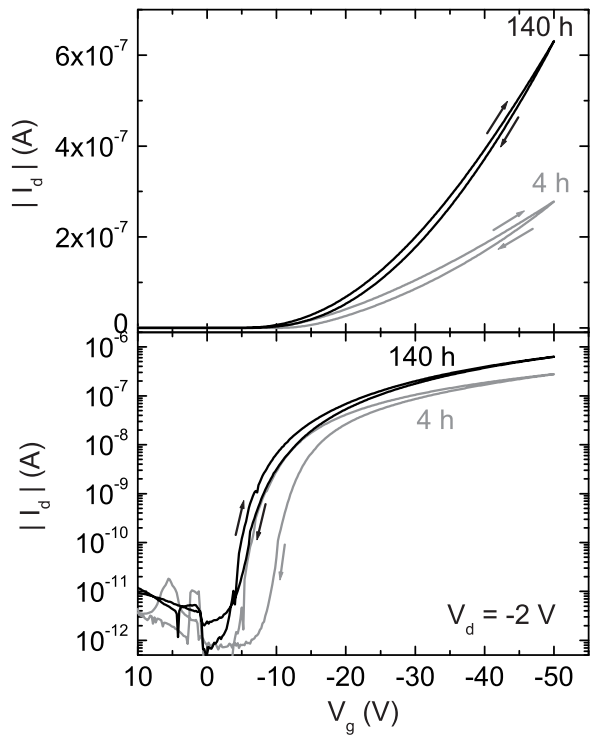}
\caption{\label{currents} Transfer characteristic from a pentacene
TFT measured 4\,h and140\,h after the completion of the pentacene
evaporation. The storage under high vacuum conditions leads to an
increased on-current and to a reduced current hysteresis.}
\end{figure}

Fig.~\ref{currents} shows the transfer characteristic from a
pentacene TFT measured 4\,h and 140\,h after the completion of the
pentacene evaporation. The device was kept at $3\times10^{-8}$\,mbar
all along. After 140\,h, the device shows an increased on-current.
In addition, the current hysteresis is reduced: at a current level
of $10^{-10}$\,A, the difference between the forward and the reverse
sweep is 3.8\,V after 4\,h and 1.2\,V after 140\,h. The subthreshold
swing is essentially unaffected by the high vacuum storage. There is
a small shift of the onset voltage to more positive voltages from
-6.4\,V after 4\,h to -4.9\,V after 140\,h.

\begin{figure}
\includegraphics[width=0.9\linewidth]{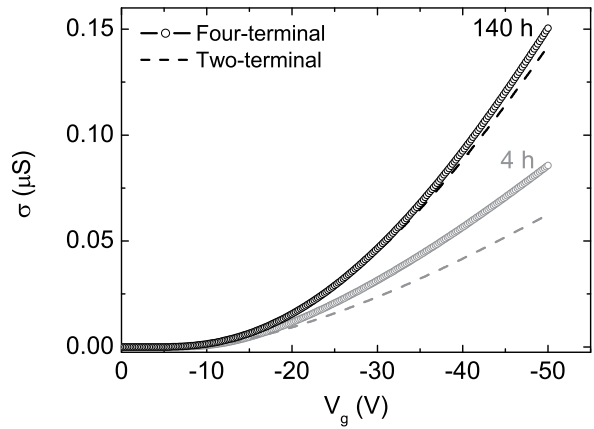}
\caption{\label{sigma} The ``four-terminal conductivity'' increases
with time. The graph shows the ``four-terminal conductivity''
(circles) after 4\,h and after 140\,h. The dashed lines indicate the
``two-terminal conductivity'' for comparison.}
\end{figure}

An increase in on-current can be due to changes of the pentacene
film and/or to a reduction of the device contact resistance. The
gated four-terminal method can disentangle the field-effect
conductivity and the device contact resistance. In Fig.~\ref{sigma}
we show the ``four-terminal conductivity'' after 4\,h and after
140\,h, as derived from the forward sweeps (Eq.~\ref{sigma2}). The
``two-terminal conductivity'' (Eq.~\ref{sigma0}) is shown for
comparison. The ``four-terminal conductivity'' is increased after
140\,h, which reveals changes of the pentacene film. The difference
between the ``four-terminal conductivity'' and the ``two-terminal
conductivity'' is reduced after 140\,h, indicative of an additional
contact resistance reduction.

\begin{figure}
\includegraphics[width=0.9\linewidth]{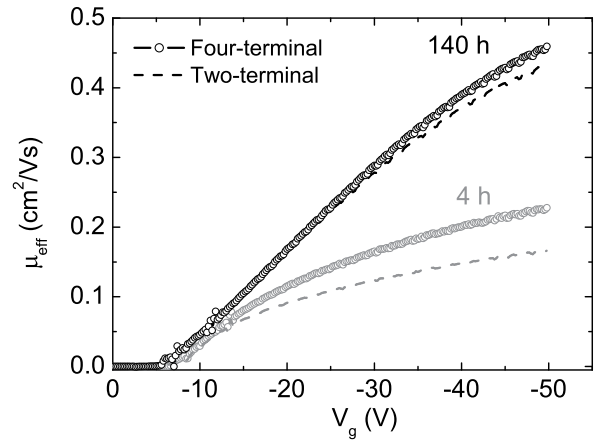}
\caption{\label{mobility} ``Four-terminal mobility'' (circles) as a
function of gate voltage. There is a significant improvement in
mobility with time. At $V_{g}\approx-50$\,V, the mobility increases
from $\mu=0.22$\,cm$^{2}$/Vs after 4\,h to $\mu=0.45$\,cm$^{2}$/Vs
after 140\,h, i.e. by a factor of two. The dashed lines show the
``two-terminal mobilities'' for comparison.}
\end{figure}

Fig.~\ref{mobility} shows the ``four-terminal mobility'' derived
with Eq.~\ref{mu} and the ``two-terminal mobility'' calculated from
Eq.~\ref{mum1}. As expected, the mobility increases monotonically
with gate voltage. When comparing both measurements, a significant
improvement in mobility can be ascertained. At $V_{g}\approx-50$\,V,
the mobility is $\mu=0.22$\,cm$^{2}$/Vs after 4\,h and
$\mu=0.45$\,cm$^{2}$/Vs after 140\,h, i.e. $\mu$ has increased by a
factor of 2.\footnote{We compare mobilities for comparable total
charge densities. Since we only observe small onset voltage shifts,
a correction of the gate voltage by the onset voltage is not
necessary.}

\begin{figure}
\includegraphics[width=0.9\linewidth]{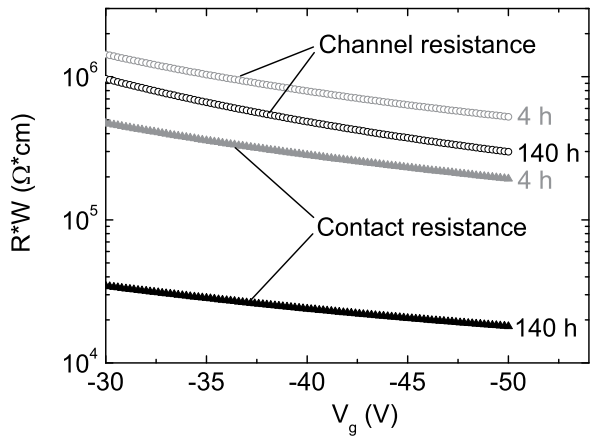}
\caption{\label{resistance} Width-normalized contact resistance
(triangles) after 4\,h and after 140\,h. The contact resistance is
drastically reduced: at $V_{g}=-50$\,V it decreases by a factor of
$\approx11$. The graph also contains the respective width-normalized
channel resistances (circles) for comparison. The contact resistance
is always lower than the channel resistance.}
\end{figure}

In Fig.~\ref{resistance} we show the width-normalized contact
resistance $R_{contact}W$ according to Eq.~\ref{rcontact} and, for
comparison, the width-normalized channel resistance $R_{channel}W$
(Eq.~\ref{rchannel}). There is a drastic reduction in contact
resistance. At $V_{g}=-50$\,V the contact resistance decreases by a
factor of $\approx11$ from $1.95\times10^{5}$\,$\Omega$cm to
$1.81\times10^{4}$\,$\Omega$cm. The channel resistance decreases by
a factor of $\approx2$. The channel resistance is always higher than
the contact resistance: at $V_{g}=-50$\,V and after 4\,h the channel
resistance is $\approx3$ times larger than the contact resistance
and after 140\,h it is $\approx17$ times the contact resistance.
Thus, the device is always dominated by the channel resistance.

\subsection{Influence on the density of states function}
\label{secdos}

\begin{figure*}
\includegraphics[width=0.8\linewidth]{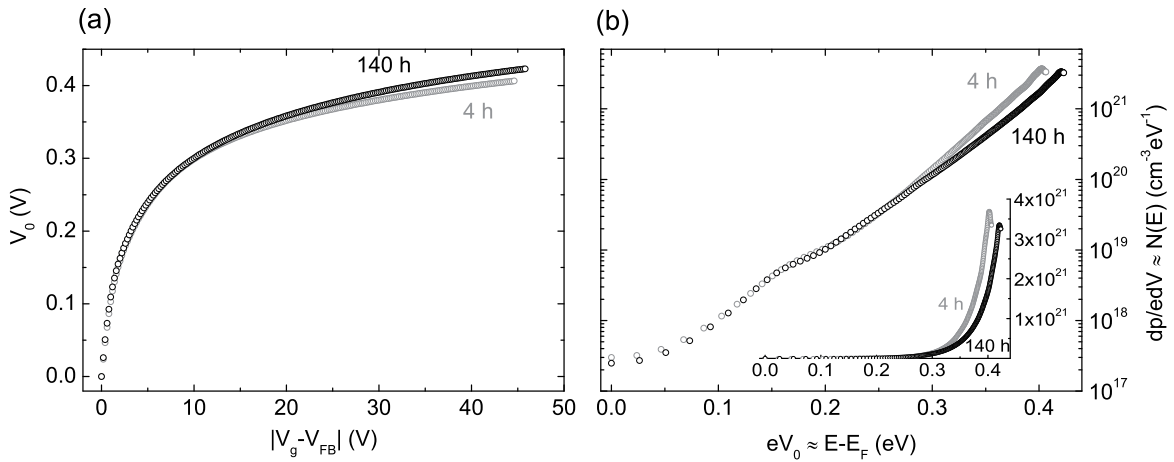}
\caption{\label{together}(a) Interface potential $V_{0}$ as a
function of gate voltage above flatband $|V_{g}-V_{FB}|$ for the
measurement after 4\,h and after 140\,h. (b) Density of trap states
as a function of energy. The main panel shows the $dp/edV$ data as a
function of the band shift at the interface $eV_{0}$ on a
logarithmic scale. The band shift at the interface is approximately
equal to the energy of the traps above the Fermi energy of the
sample, i.e. $eV_{0}\approx E-E_{F}$. The quantity $dp/edV$ is a
good approximation of the density of trap states $N(E)$. The high
vacuum storage leads to a reduced density of traps close to the
valence band edge. The inset show the trap densities on a linear
scale.}
\end{figure*}

In Sec.~\ref{evstart} and \ref{evend} we show, that the performance
improvement is not due to doping by a residual gas but due to a
healing of defects at room temperature. To investigate the energetic
position of these defects, we have applied the scheme described in
Sec.~\ref{advanced}. The first step was to calculate the interface
potential $V_{0}$ as a function of gate voltage with
Eq.~\ref{implicit}. We have assumed a dielectric constant of
$\epsilon_{i}=3.9$ for SiO$_{2}$ and $\epsilon_{s}=3.0$ for
pentacene. Fig.~\ref{together}(a) shows the result of the extraction
for the measurement after 4\,h and after 140\,h. In a second step,
the density of states was calculated by two numerical
differentiations of $V_{0}(V_{g})$ according to Eq.~\ref{total2} and
Eq.~\ref{total3}. Some degree of data smoothing was applied in order
to obtain a reasonably smooth density of states function.
Fig.~\ref{together}(b) shows the final result for the measurement
after 4\,h and after 140\,h. A step width of 0.2\,V in the gate
voltage sweeps leads to a good resolution of the deep states.

The Fermi energy of the sample coincides with the zero point of the
energy scale in Fig.~\ref{together}(b). At high gate voltages, the
valence band edge at the interface is close to the Fermi level.
Consequently, the Fermi level is approximately $0.45$\,eV from the
valence band edge at flatband. For comparison, the bandgap of
pentacene is 2.24-2.5\,eV.\cite{GundlachDJ2006} The interface
potential in Fig.~\ref{together}(a) reflects the spectral density of
trap states. At low gate voltages bands bend easily, which indicates
a low trap density. At high gate voltages, however, band bending is
more difficult and this is indicative of a high trap density very
close to the valence band edge.

The energetics in Fig.~\ref{together}(a) are in good agreement with
activation energies $E_{a}(V_{g})$ from temperature dependent
measurements of pentacene-based
transistors.\cite{VissenbergMCJM1998, LangDV2004, DeAngelisF22006}
In these studies the activation energy is derived from
$\sigma(V_{g})\propto\exp(-E_{a}(V_{g})/kT)$ by means of Arrhenius
plots. Assuming the abrupt approximation, Eq.~\ref{sigma3} can be
approximated as
\begin{eqnarray}
\sigma&=&e\mu_{0}N_{V}\exp\bigg(-\frac{E_{V}-E_{F}}{kT}\bigg)\int_{0}^{d}\exp\bigg(\frac{eV(x)}{kT}\bigg)dx\nonumber\\
& \approx
&e\mu_{0}N_{V}\lambda(V_{g})\exp\bigg(-\frac{E_{V}-E_{F}-eV_{0}}{kT}\bigg)
\end{eqnarray}
and we have $E_{a}(V_{g})\approx E_{V}-E_{F}-eV_{0}$. The activation
energy from the Arrhenius plots is approximately equal to the
energetic difference between the Fermi level and the valence band at
the interface. The activation energy is found to be $0.3-0.6$\,eV
near the onset voltage of the devices and $0.15-0$\,eV when the
devices are turned on completely. $E_{a}(V_{g})$ shows a similar
functional dependence on gate voltage as $V_{0}(V_{g})$ in
Fig.~\ref{together}(a).\cite{VissenbergMCJM1998, LangDV2004,
DeAngelisF22006}

From Fig.~\ref{together}(b) it is clear that it is the shallow traps
with energies approximately $0.15$\,eV from the valence band edge
which are reduced by the high vacuum storage. It is the density of
these states that influences the value of the field-effect mobility
$\mu_{eff}$. A relatively small reduction leads to a significant
improvement in field-effect mobility. The traps which are deeper in
energy are essentially unaffected, resulting in an almost identical
subthreshold swing of the transfer characteristics.

Both after 4\,h and after 140\,h, the density of states function can
reasonably well be approximated by a single exponential function
\begin{equation}\label{expo}
N(E)=N_{0}\exp\bigg(\frac{E}{E_{0}}\bigg).
\end{equation}
It is however slightly steeper than exponential. Fitting the curves
in Fig.~\ref{together}(b) for $eV_{0}\geq0.25$\,eV to Eq.~\ref{expo}
yields the parameters $E_{0}=32$\,meV for the measurement after 4\,h
and $E_{0}=37$\,meV for the measurement after 140\,h. This is in
good agreement with results for pentacene TFTs obtained with a
device simulation program.\cite{OberhoffD2007, PernstichKP2007} In
Ref.~\onlinecite{OberhoffD2007}, characteristic slopes of
$E_{0}=34$\,meV and 37\,meV are specified and in
Ref.~\onlinecite{PernstichKP2007} a slope of $E_{0}=32$\,meV was
determined.

\subsection{Comparison of several experiments}

\begin{figure}
\includegraphics[width=0.8\linewidth]{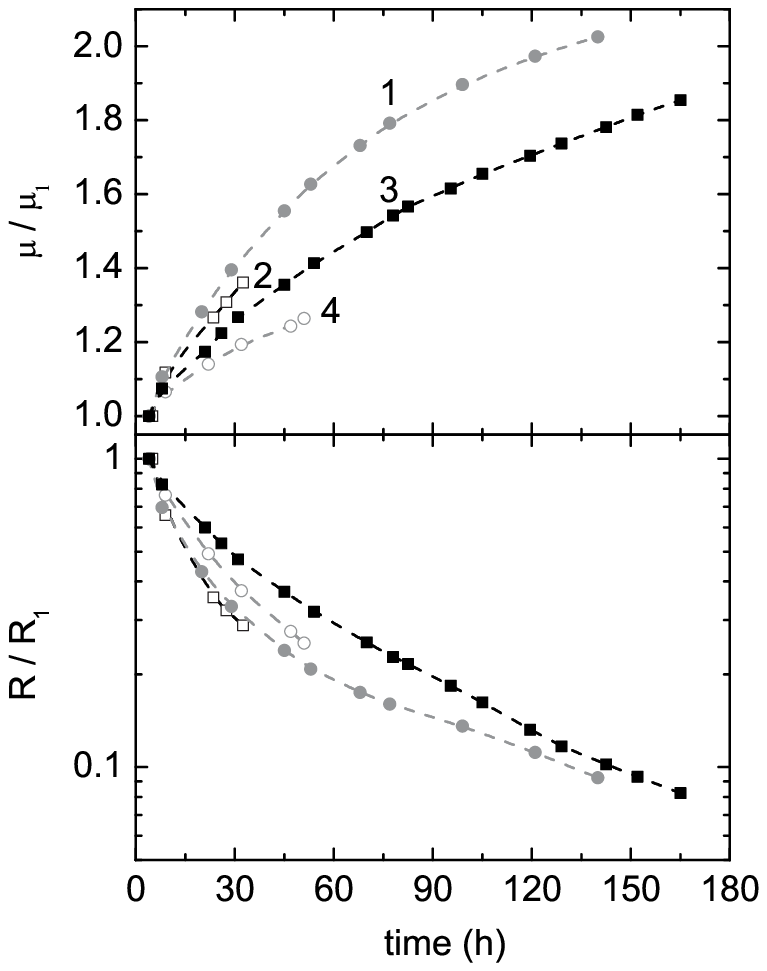}
\caption{\label{nointerference} Upper panel: ``four-terminal
mobility'' at $V_{g}\approx-50\,V$ normalized by the mobility
$\mu_{1}$ of the initial measurement for four different runs. The
lower panel shows, on a logarithmic scale, the respective values for
the contact resistance $R$ at $V_{g}=-50\,V$ relative to the contact
resistance $R_{1}$ from the initial measurement. The field-effect
mobility increases with time and the contact resistance reduces with
time in all experiments. There is some variation in the rate of the
effects.}
\end{figure}

The effects described above, i.e. a significant increase in the
``four-terminal mobility'', a drastic reduction in the contact
resistance and a reduction in the current hysteresis, have been
observed in all experiments. Fig.~\ref{nointerference} shows the
evolution of the ``four-terminal mobility'' and the contact
resistance with time for four different runs. The values are for
$V_{g}\approx-50\,V$ and are normalized by the value obtained after
$\approx4$\,h.\footnote{The value for the contact resistance was
taken at $V_{g}=-50\,V$. For the mobility, an average in the range
$V_{g}=-45$\,V to $-50$\,V was taken since in some cases the
derivative of the drain current was more noisy.} A time span of at
least 4\,h was allowed in between subsequent measurements. The
absolute values for the mobility and the contact resistance from the
initial measurement are summarized in Table~\ref{table1}.
\begin{table}
\caption{\label{table1} ``Four-terminal mobility'' $\mu_{1}$, at
$V_{g}\approx-50$\,V, width-normalized contact resistance $R_{1}W$
at $V_{g}=-50$\,V and onset voltage $V_{on}$ from the initial
measurements of different experiments. The mobility increased by as
much as a factor of 2 in the course of an experiment such that
mobilities up to 0.45\,cm$^{2}$/Vs were achieved.}
\begin{ruledtabular}
\begin{tabular}{lccr}
Run & $\mu_{1}$ [cm$^{2}$/Vs]& $R_{1}W$ [$\Omega$cm]& $V_{on}$ [V] \\
\hline 1 & 0.22 & 1.95$\times10^{5}$
& -6.4\\
2 & 0.24 & 1.10$\times10^{5}$ & -4.1\\
3 & 0.14 & 3.03$\times10^{5}$ & -5.1\\
4 & 0.14 & 3.35$\times10^{5}$ & -6.4\\ \hline
Oxygen & 0.12 & 3.98$\times10^{5}$ & -6.2\\
Nitrogen & 0.10 & 4.70$\times10^{5}$ & -7.0\\
Annealing & 0.10 & 4.55$\times10^{5}$ & -5.7\\
\end{tabular}
\end{ruledtabular}
\end{table}

Table~\ref{table1} also contains the respective onset voltages.
Initially, the onset voltage is between $-4.1$\,V and $-6.4$\,V and
shifts in all cases by less than 2.5\,V during the course of an
experiment. The reduction of the current hysteresis takes place in
the early stages of the experiments. In all four runs, the current
hysteresis is significantly reduced after the first $\approx24$\,h
to $1.0-1.5$\,V at a current level of $10^{-10}$\,A. Subsequently,
there is only a small further reduction of the current hysteresis.

We now proceed by providing experimental evidence that the
performance improvement is not due to doping of the pentacene thin
films by residual oxygen or nitrogen and show that the performance
improvement is a thermally promoted process.

\subsection{Influence of oxygen and nitrogen} \label{evstart}

Even at a pressure of order $10^{-8}$\,mbar the time for the
formation of a monolayer of residual gas molecules is less than ten
minutes.\cite{RothA1990} In the case of semiconducting polymers
experimental evidence indicates that doping leads to an increased
field-effect mobility.\cite{BrownAR1997} This may be understood if
one assumes the charge transport to take place by variable-range
hopping and the doping to increase the density of states/hopping
sites.\cite{BrownAR1997} In crystalline small molecule organic
semiconductors, however, charge transport in extended states has to
be considered. Doping raises the Fermi level of the sample. This
results in an increase of the (flatband) conductivity, since the
density of free holes is increased (Eq.~\ref{sigmanull}). When the
device is turned on, however, the current flow is due to the gate
induced holes and the holes from a chemical doping are negligible in
Eq.~\ref{mu0}. Thus it would not immediately be obvious how chemical
doping would increase the field-effect mobility in pentacene.

Both oxygen and nitrogen have been reported to have the capability
of doping pentacene. Gas exposure is found to lead to a shift of the
transistor transfer characteristic to more positive voltages which
corresponds to a shift of the Fermi level.\cite{OgawaS2005,
KnippD2007} In other studies, an increase in conductivity of gas
exposed pentacene thin-films or single crystals was observed in
two-terminal measurements without gate-field induced charge. The
increased conductivity is ascribed to an increased charge carrier
density caused by doping.\cite{KurodaH1961, JurchescuOD2005,
ParisseP2006} There seems to be no evidence that doping leads to an
increased effective mobility in pentacene. Recent measurements on
rubrene single crystals also show doping not to increase the
field-effect mobility.\cite{SoWY2007}

The issue of reversibility and the importance of light for the
doping is not completely clear. Doping is observed when pentacene
thin films are exposed to relatively high partial pressures
(0.01\,atm-1\,atm) of oxygen or nitrogen in the presence of
light.\cite{KurodaH1961, OgawaS2005, ParisseP2006} The doping effect
by oxygen is reported to be negligible in the absence of
light.\cite{KurodaH1961, OgawaS2005} Ultraviolet photoelectron
spectroscopy (UPS) before and after exposure to
$5\times10^{-6}$\,mbar oxygen could not detect a lasting effect on
the position of the energy levels.\cite{VollmerA2005} According to
Ref.~\onlinecite{KnippD2007}, exposing pentacene TFTs to an oxygen
partial pressure of $10^{-5}$\,mbar leads to a doping of the films.

While the mobility increase in crystalline/polycrystalline samples
cannot be easily linked to chemical doping, a reduction of the
contact resistance might be reconciled with doping. From inorganic
semiconductor physics it is well known that doping close to a
contact reduces the contact resistance. Moreover, it has been shown
with UPS that oxygen, at a low partial pressure, can lead to a
lowering of the injection barrier for holes at an Au/pentacene
interface.\cite{VollmerA2005}

\begin{figure}
\includegraphics[width=0.88\linewidth]{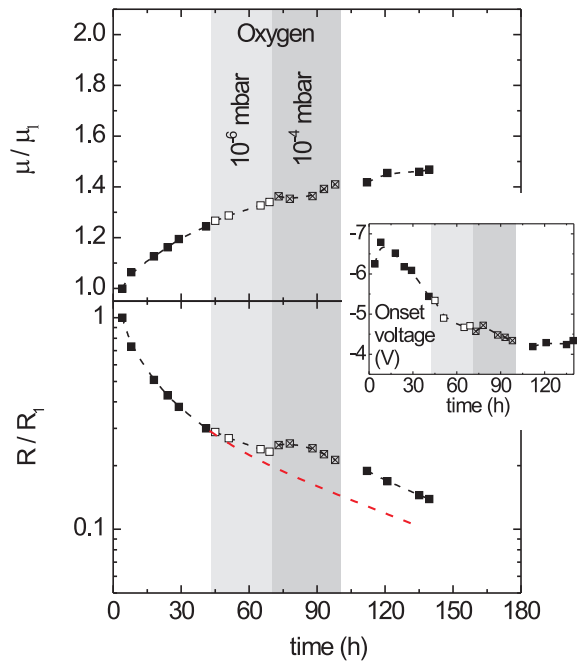}
\caption{\label{oxygengrey} Effect of oxygen exposure: the graph
shows the ``four-terminal mobility'' at $V_{g}\approx-50\,V$ (upper
panel) and the contact resistance at $V_{g}=-50\,V$ (lower panel) as
a function of time. The dashed red line is an estimate of the
time-dependence of the contact resistance without oxygen exposure.
The drastic increase in oxygen partial pressure neither leads to a
more rapid increase in mobility nor to a more rapid decrease in
contact resistance. On the contrary, the oxygen exposure slows down
the decrease in contact resistance. In the inset we show the device
onset voltage as a function of time. The increase in oxygen partial
pressure does not lead to a sudden shift of the onset voltage which
would be indicative of doping.}
\end{figure}

The results of our studies concerning the effect of gas exposure on
the field-effect mobility and on the contact resistance are
discussed in the following. If e.g. oxygen was responsible for the
increase in mobility and/or reduction in contact resistance, an
increase in the oxygen partial pressure should accelerate the rate
of the respective process. Fig.~\ref{oxygengrey} shows the influence
of oxygen on the ``four-terminal mobility'' and on the contact
resistance (see Table.~\ref{table1} for the initial parameters).
After 41\,h, the oxygen partial pressure was raised from a total
turbo-pumped background pressure of order $10^{-8}$\,mbar to
$10^{-6}$\,mbar. After 69\,h, the partial pressure was increased to
$10^{-4}$\,mbar and eventually, after 98\,h, the chamber was
re-evacuated. The increase in oxygen partial pressure does not
accelerate the gradual increase in mobility or the reduction in
contact resistance. The dashed red line in Fig.~\ref{oxygengrey} is
an estimate for the time-dependence of the contact resistance if the
sample had not been exposed to oxygen. Consequently, the increase in
oxygen partial pressure even leads to a delay of the reduction in
contact resistance. It is noteworthy that neither the small current
hysteresis nor the subthreshold swing of the devices are effected by
the oxygen exposure. Since oxygen at a partial pressure of
$10^{-6}$\,mbar for 28\,h and at a partial pressure of
$10^{-4}$\,mbar for 29\,h appears not to accelerate the development
of the device parameters with time one would conclude that oxygen at
the base pressure is not responsible for the performance
improvement.

The inset in Fig.~\ref{oxygengrey} shows the onset voltage for each
of the measurements. The overall shift of the onset voltage is small
and smooth ($\approx2.6$\,V over $\approx140$\,h). In particular,
there is no sudden shift of the transfer characteristic after an
increase in oxygen partial pressure. Therefore, there is no evidence
that oxygen at partial pressures $\leq10^{-4}$\,mbar leads to a
doping of the films.

\begin{figure}
\includegraphics[width=0.88\linewidth]{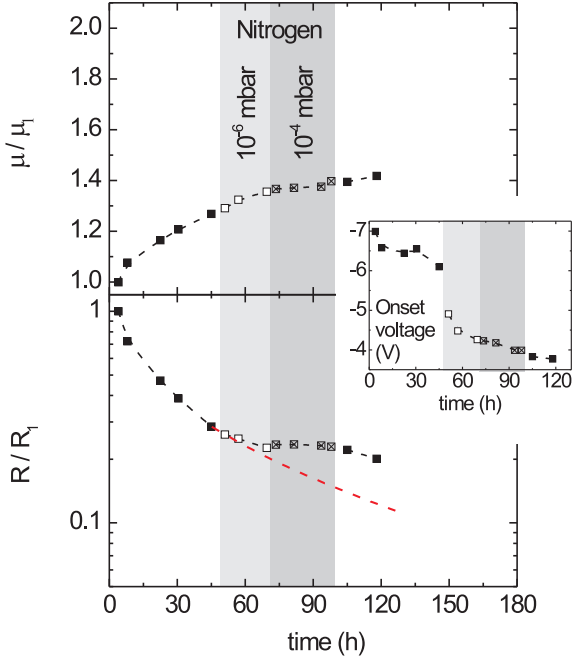}
\caption{\label{nitrogengrey} Effect of nitrogen exposure: a
nitrogen partial pressure of $10^{-4}$\,mbar leads to a slowing of
the increase in mobility and decrease in contact resistance. The
dashed red line is an estimate for the time-dependence of the
contact resistance if the sample had not been exposed to nitrogen.
The inset shows the device onset voltage. When the nitrogen partial
pressure is increased to $10^{-6}$\,mbar, there is a sudden shift of
the onset voltage which is indicative of doping.}
\end{figure}

Fig.~\ref{nitrogengrey} shows the analogous experiment with
nitrogen. After 50\,h the nitrogen partial pressure was increased to
$10^{-6}$\,mbar, followed by an increase to $10^{-4}$\,mbar after
69.5\,h. After 98\,h, the system was pumped down to the base
pressure of order $10^{-8}$\,mbar. The increase in nitrogen partial
pressure does not accelerate the increase in mobility or the
reduction in contact resistance. Similarly to the oxygen experiment
we have estimated the time-dependence of the contact resistance if
the sample had not been exposed to nitrogen (dashed red line in
Fig.~\ref{nitrogengrey}). Clearly, the nitrogen exposure leads to a
slowing down of the decrease in contact resistance. The nitrogen
exposure leaves the small current hysteresis and the subthreshold
swing unaffected. Residual nitrogen is not responsible for the
performance improvement.

The inset in Fig.~\ref{nitrogengrey} shows the device onset voltage
as a function of time. When the nitrogen partial pressure is
increased to $10^{-6}$\,mbar there is a sudden shift of the onset
voltage to more positive voltages by 1.2\,V. We take this as
evidence for doping by nitrogen at a partial pressure of
$10^{-6}$\,mbar. When the pressure is increased to $10^{-4}$\,mbar,
there is no additional marked shift of the onset voltage.

\subsection{Annealing at slightly elevated temperatures} \label{evend}

\begin{figure}
\includegraphics[width=0.8\linewidth]{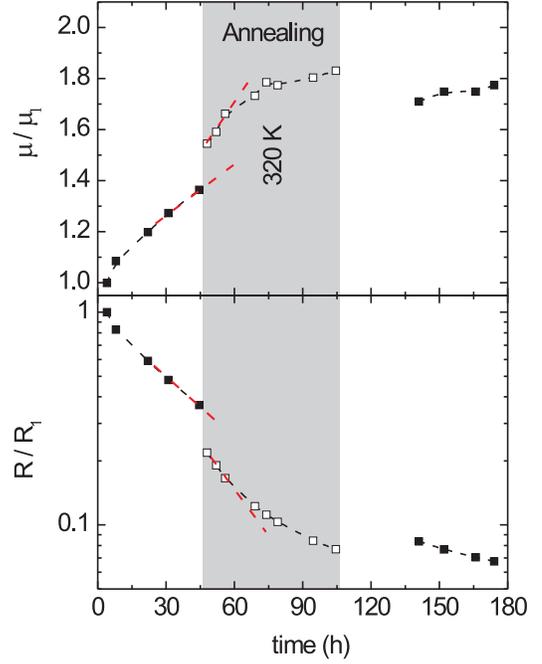}
\caption{\label{annealinggrey} Effect of annealing at 320\,K: the
increased temperature accelerates the increase in mobility and the
decrease in contact resistance as indicated by the dashed red
lines.}
\end{figure}

Annealing pentacene thin films at moderate temperatures (e.g.
50$^{\circ}$\,C) results in an improved crystallinity of the films
as seen by XRD.\cite{KomodaT2002, YeR2003, KangS2004} When
concentrating on the channel region adjacent to the gate dielectric
in a field-effect transistor, the annealing is found to leave the
mobility unchanged (Ref.~\onlinecite{KomodaT2002}) or to lead to an
increased mobility (Ref.~\onlinecite{KangS2004}).
Since the interaction between the pentacene molecules is of the weak
Van der Waals type, it is not too surprising that annealing at
moderate temperatures leads to an improved crystallinity of the
films. Moreover, it seems plausible that structural defects can be
``annealed'' even at room temperature, which will account for the
effects we observe.
In this scenario we would expect an acceleration of the performance
improvement at higher temperatures. Fig.~\ref{annealinggrey} shows
the influence of such an annealing at 320\,K (see Table~\ref{table1}
for the initial parameters). After 44.5\,h, the sample temperature
was slowly raised from RT to 320\,K at a rate of
$0.2$\,$^{\circ}$/min. Transfer characteristics were measured after
the temperature of 320\,K had been reached. After 104.5\,h, the
heating was switched off, followed by a slow cooling of the sample
to room temperature. Two effects can be discerned in the annealing
process. First, the overall level of the mobility and the contact
resistance are affected by the increase in temperature since both
quantities depend on temperature. Second, and more significant in
the present context, is the rate of change: after raising the
temperature, the increase in mobility and the decrease in contact
resistance is accelerated significantly. This is indicated by the
dashed red lines in Fig.~\ref{annealinggrey}.

In an additional experiment (not shown), the sample temperature was
raised from room temperature to 310\,K after 45.5\,h and was
increased to 320\,K after 74\,h. Even at 310\,K, which is not much
above room temperature ($\approx297\,K$), the increase in mobility
and the decrease in contact resistance is accelerated noticeably.

All these results taken together clearly indicate an improvement of
the electronic parameters driven by a thermally promoted process,
and not by chemical doping. We suggest this process to be a healing
of structural defects. We now discuss the microscopic origin of the
relevant traps and suggest how a reduced trap density can lead to a
reduced contact resistance.

\section{Discussion}

\subsection{Defect healing at room temperature}

Pentacene thin-films on Si/SiO$_{2}$ substrates are known to have a
layered structure and the layers are parallel to the substrate.
Within these layers, the molecules are arranged in a herringbone
pattern and are oriented almost perpendicular with respect to the
substrate. It has been shown by high impedance STM that some of the
pentacene molecules in the layers are displaced along their long
molecular axis, while the two-dimensional packing is not
disturbed.\cite{KangJH2005} With electronic structural calculations
it could be shown that the displaced molecules result in traps
$\leq0.1$\,eV from the valence band edge.\cite{KangJH2005}

In Sec.~\ref{secdos} we have shown that only the shallow traps
$\leq0.15$\,eV from the valence band edge are reduced during the
high vacuum storage at room temperature. We suggest that a major
cause of the shallow traps in pentacene thin films are pentacene
molecules within the grains that are slightly misplaced, i.e.
various types of structural point defects. Some of these defects are
in a metastable state before relaxation. They require only a small
amount of energy in order to align which can be provided by the
thermal energy at room temperature. This is a direct manifestation
of the weak intermolecular interaction which is characteristic of
small molecule organic semiconductors. The ``annealing'' of shallow
traps at room temperature can easily explain the increase in the
effective field-effect mobility.

\subsection{Defects and contact resistance}

\begin{figure}
\includegraphics[width=0.6\linewidth]{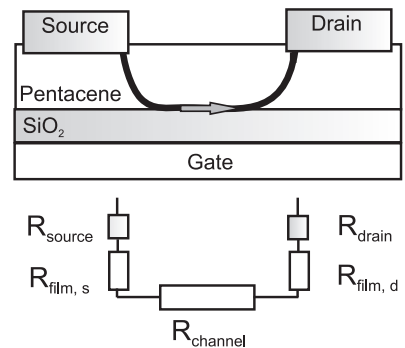}
\caption{\label{transistor} Sketch of the different contributions to
the total contact resistance. The holes must overcome a barrier
associated to the gold/pentacene interface. The resistance of the
50\,nm thick pentacene film adds to the total contact resistance
both at the source and at the drain.}
\end{figure}

In a simplistic view, the contact resistance is given by the
energetic difference between the work function of the metal and the
ionization energy of the pentacene. In reality, however, a clear
correlation between the metal work function and the contact
resistance is often not observed.\cite{GundlachDJ2006} Interface
states can significantly affect the energetics at the
metal-pentacene interface.\cite{PesaventoPV2004} In a top contact
device, also the film resistance should contribute to the contact
resistance. The film resistance in pentacene devices with gold top
contacts has even been suggested to dominate the contact
resistance.\cite{PesaventoPV2006} The situation is illustrated in
Fig.~\ref{transistor}: the hole injection at the source/pentacene
interface is good, but the holes must cross the pentacene film in
order to reach the channel at the insulator-semiconductor interface.
The intrinsic mobility perpendicular to the molecular layers is
lower than parallel to the layers. Combined with high trap
densities, this can result in a large resistance between the gold
electrodes and the transistor channel. If the contact resistance is
dominated by the film resistance, a reduction in the density of hole
traps is therefore expected to lead to a reduced contact resistance.

\section{Summary and conclusions}

Pentacene thin-film transistors were made by thermal evaporation,
employing a high precision in situ mask alignment mechanism. The
devices were characterized electrically by gated four-terminal
measurements without breaking the high vacuum (base pressure of
order 10$^{-8}$\,mbar). Under the high vacuum conditions, the device
performance is found to improve with time. Within approximately one
week, the contact-corrected field-effect mobility improves by a
factor of up to two and the device contact resistance typically
decreases by more than an order of magnitude. In addition, the
current hysteresis reduces significantly. We have shown that an
increased partial pressure of oxygen or nitrogen does not accelerate
the performance improvement. On the contrary, the gas exposure
delays the decrease in contact resistance. Nitrogen was found to
dope pentacene thin films at partial pressures as low as
10$^{-6}$\,mbar. Annealing at a slightly elevated temperature (e.g.
320\,K), on the contrary, leads to an acceleration of the
performance improvement.

Some defects within the pentacene ``anneal'' even at room
temperature. This is a peculiarity of the physics of organic
semiconductors, which is governed by the weak Van der Waals type
interaction between the molecules. We have derived the spectral
density of trap states from the ``four-terminal conductivity''. The
calculations show shallow traps $\leq0.15$\,eV from the valence band
edge to be significantly affected by the defect healing. We suggest
these traps to originate from structural point defects, i.e.
slightly misaligned molecules within the grains of the
polycrystalline film. The effective field-effect mobility critically
depends on the number of these shallow traps and a relatively small
reduction results in a significant improvement of the mobility. The
contact resistance is likely to be dominated by the film resistance
and also depends on the active traps within the film.

The method to calculate the spectral density of traps is a powerful
tool to further elucidate the origin of trap states in organic
semiconductors, provided that contact effects are properly taken
into account. It is particularly suitable to study metastable
defects in organic semiconductors, because the density of states
function can be derived from a single transfer characteristic in an
unambiguous fashion and with a minimal set of simplifying
assumptions.

\begin{acknowledgments}
The authors thank K.~P. Pernstich, D.~J. Gundlach and S. Haas for
their help with the experimental setup and for valuable discussions.
\end{acknowledgments}


\newpage


\end{document}